\documentclass[12pt, letterpaper]{article}
\usepackage[top=1in, bottom=1.5in, left=1in, right=1in]{geometry}
\usepackage[utf8]{inputenc}
\usepackage{booktabs}
\usepackage{hyperref}
\usepackage{natbib}
\usepackage{multirow}
\usepackage{array}
\usepackage{graphicx}

\usepackage{rotating}

\usepackage{authblk}
\title{TDEs with LSST}
\author[1]{Katja Bricman\thanks{\href{mailto:katja.bricman@ung.si}{katja.bricman@ung.si}}}
\author[1]{Andreja Gomboc}
\author[ ]{\\ \small{with the support of the LSST Transient and Variable Stars Collaboration}}
\affil[1]{ \small Center for Astrophysics and Cosmology,
University of Nova Gorica, Vipavska 11c, SI-5270 Ajdovščina, Slovenia}
\date{December 2018}

\begin{document}

\maketitle

\begin{abstract}
We investigate the prospects of observing Tidal Disruption Events (TDEs) with different LSST cadences proposed with the White paper call. We study their detection rates, the quality of their light curves and discuss which cadences sample TDEs better before or during the peak of the light curve. We suggest some constraints on the observing cadences that we believe will increase the number of reliable classifications of TDEs, in particular a WFD survey with 2 visits in different filters every night or at least every second night, observing the extra-galactic sky.
\end{abstract}

\section{White Paper Information}
\begin{enumerate} 
\item {\bf Science Category:} Transeints and variable stars
\item {\bf Survey Type Category:} Wide-fast-deep survey
\item {\bf Observing Strategy Category:} 
a specific observing strategy to enable specific time domain science, 
that is relatively agnostic to where the telescope is pointed (e.g., a science case enabled 
by relatively deep precise time-resolved multi-color photometry).
\end{enumerate}

\clearpage

\section{Scientific Motivation}
When a star in a nucleus of a galaxy gets scattered into the vicinity of the
supermassive black hole (SMBH) in the core of the galaxy, the star can be
torn apart by black hole's strong tidal forces \citep{Rees:1988bf, 
Phinney1989, Evans:1989qe}. The process, known as a Tidal Disruption Event
(TDE), can emit a bright flare of light, which slowly decays with time on time
scales from months to years.

TDEs are very promising probes for studies of quiescent black holes, which form 
the large majority of SMBHs found in nuclei of
most galaxies in the universe. Observed light curves of TDEs hold information
on the SMBH mass involved, as well as on other parameters, such as the
mass and the radius of the disrupted star, the distance from the black hole 
at which the star is disrupted, and others \citep{Kochanek:1993cm, Gomboc:2005wu, 
Lodato:2008fr, Strubbe:2009qs, Lodato:2010xs, Guillochon:2012uc, Mockler:2018xne}

\begin{figure}[h]
\centering
\includegraphics[width=\linewidth]{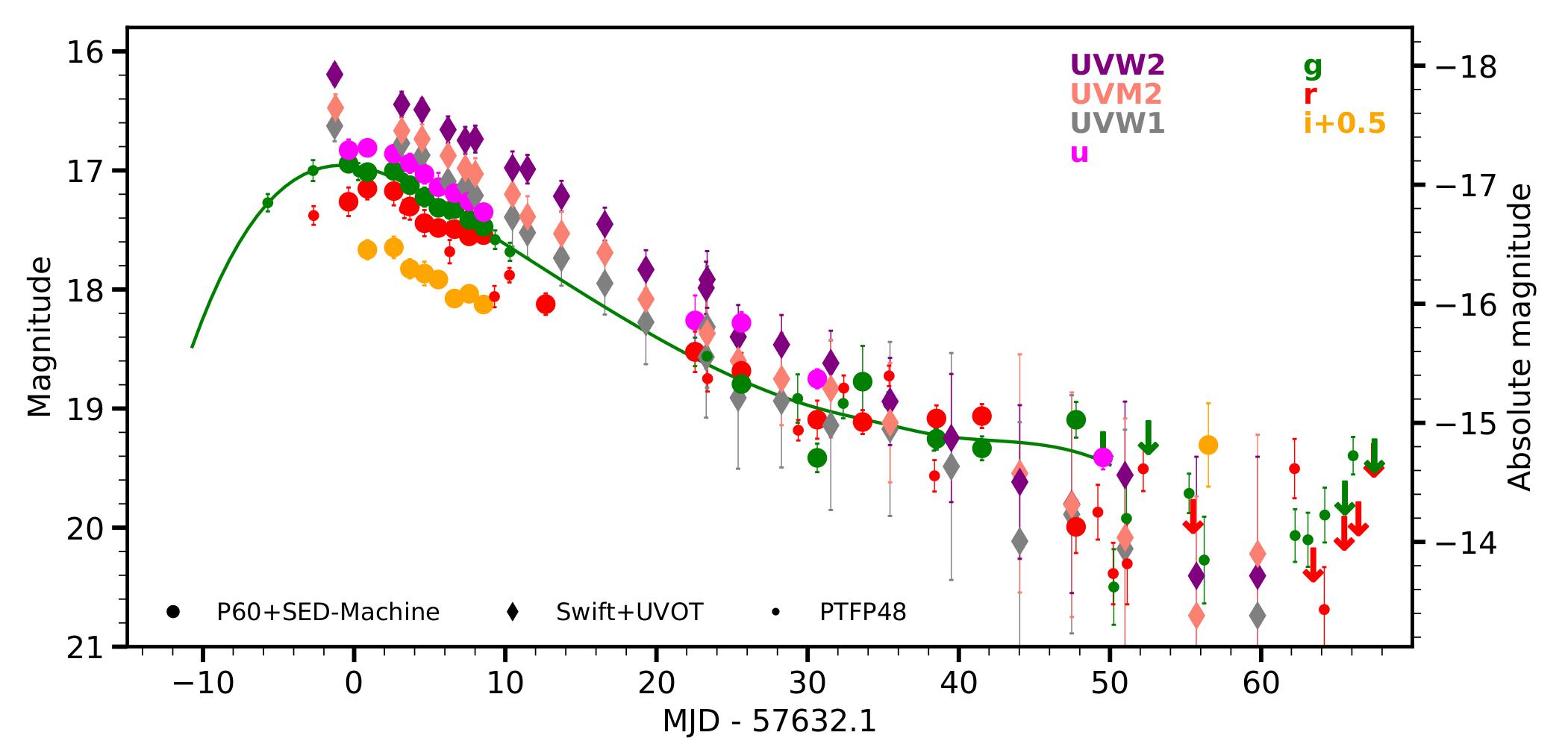}
\caption{Light curve of a TDE discovered with iPTF in 2016 \citep{Blagorodnova:2017gzv}.}
\label{iPTF-16fnl}
\end{figure}

TDEs are rare transients, with only a few tens of candidates discovered 
so far \citep[e.g. ][]{vanVelzen:2010jp, Gezari:2012sa, Arcavi:2014xxx, 
Chornock:2013jta, Holoien:2014jha, Holoien:2015pza, Holoien:2016uaf, 
Leloudas:2016rmh, Wyrzykowski:2016acu, Blagorodnova:2017gzv}. A
typical light curve is shown in Figure \ref{iPTF-16fnl}. The rate
of TDEs was estimated to be between $10^{-4}$ and $10^{-5}$ per galaxy
per year \citep{Magorrian:1999vm, Wang:2003ny}. Due to the small rate,
large surveys which are constantly observing billions of galaxies, such as 
the LSST, are more likely to catch these events and will substantially enlarge the 
observed TDE sample.

TDEs have been observed in optical wavelengths and in X-rays. 
It was recently suggested, that the observed emission depends on the 
inclination of the star-black hole orbit as viewed from Earth \citep{Dai:2018jbr}.
For the LSST only TDEs observed in optical wavelengths will be relevant,
however multi-wavelength follow-up and spectroscopic observations 
will be needed for a firm classification and for a better understanding
of the physical processes involved.

The main sources of contamination are type Ia Supernovae, which can 
exhibit a similar behavior in the light curve, especially near the peak. A few
features which can help us distinguish TDEs from SNe are their colors 
(they should be quite blue, with \emph{g-r} $\approx$ -1), their
temperature evolution, the distance from the galactic center, and spectra.

LSST should substantially enlarge the sample of observed TDEs. Estimations
have been done in the past, that the LSST should discover 40000 new TDEs
in 10 years \citep{vanVelzen:2010jp}. In \cite{Abell:2009aa} this number was 
estimated to be 60000, based on the universal TDE rates from 
\cite{Rau:2009yx}. The observations of TDEs are important for studies of 
SMBHs and stellar dynamics in central regions of galaxies, for determining
more accurately their rates and observational signatures, for exploring 
the mass distribution of black holes (in particular on their low-mass end), 
accretion physics, and jet formation.

\vspace{.6in}

\section{Technical Description}

\subsection{High-level description}
Ideally, we would only have the WFD survey, with two visits per night or
two visits every second night, where in each subsequent visit during the same night
the filters would be changed. This would help us to more accurately determine 
the color of TDEs and their temperature evolution (main characteristics,
which can distinguish them from SNe Ia). 
\vspace{.3in}

\subsection{Footprint -- pointings, regions and/or constraints}
TDEs are extra-galactic sources, therefore, due to the Galactic dust absorption,
we do not expect to detect them in the optical in the direction of Galactic plane. 
Hence, the region between galactic latitudes $-15^\circ < b < +15^\circ$ should be avoided.

\subsection{Image quality}
No constraints.

\subsection{Individual image depth and/or sky brightness}
No constraints.

\subsection{Co-added image depth and/or total number of visits}
No constraints.

\subsection{Number of visits within a night}
Ideally 2 visits in different filters per night or 2 visits in different filters every second night. 

\subsection{Distribution of visits over time}
There are no special constraints on the timing of visits within a night, 
between nights, between seasons or between years.
2 visits per night would be ideal. 2 visits every second night could also work. 
Although, concerning only the rates of discovered events, as shown later on, 
there are no large discrepancies between the single visit per night cadence 
(\texttt{Colossus\_2667}) and other cadences. 

\subsection{Filter choice}
Change of filters between subsequent visits in a night.

\subsection{Exposure constraints}
The 15 second exposure should work, it does not seem that there are any
significant changes in the detection rate if the exposure is 20 seconds long.

\subsection{Other constraints}
No other constraints.

\subsection{Estimated time requirement}
TDEs are expected to be discovered within the WFD survey.

\vspace{.3in}

\begin{table}[ht]
    \centering
    \begin{tabular}{l|l|l|l}
        \toprule
        Properties & Importance \hspace{.3in} \\
        \midrule
        Image quality & 2    \\
        Sky brightness & 3 \\
        Individual image depth &3   \\
        Co-added image depth &   3\\
        Number of exposures in a visit   & 1  \\
        Number of visits (in a night)  &  1 \\ 
        Total number of visits & 1 \\
        Time between visits (in a night) & 1 \\
        Time between visits (between nights)  & 1  \\
        Long-term gaps between visits &1 \\
        Other (please add other constraints as needed) & change filters between subsequent visits \\
        \bottomrule
    \end{tabular}
    \caption{{\bf Constraint Rankings:} Summary of the relative importance of various survey strategy constraints. The importance of each of these considerations is ranked as 1=very important, 2=somewhat important, 3=not important.}
        \label{tab:obs_constraints} 
\end{table}

\subsection{Technical trades}
\begin{enumerate}
    \item \textit{What is the effect of a trade-off between your requested survey footprint (area) and requested co-added depth or number of visits?}\\
The survey area and co-added depths remain unchanged.
    \item \textit{If not requesting a specific timing of visits, what is the effect of a trade-off between the uniformity of observations and the frequency of observations in time? e.g. a `rolling cadence' increases the frequency of visits during a short time period at the cost of fewer visits the rest of the time, making the overall sampling less uniform.}\\
Rolling cadence is disfavored, since it yields less events. However, in case
of a rolling cadence, the light curves are sampled more frequently, and the number 
of TDEs discovered in the pre-peak phase is larger, but both improve only by a factor 
of at most 2 with respect to other (non-rolling) cadences.
    \item \textit{What is the effect of a trade-off on the exposure time and number of visits (e.g. increasing the individual image depth but decreasing the overall number of visits)?}\\
 It seems there are no benefits if the exposure time is longer.
    \item \textit{What is the effect of a trade-off between uniformity in number of visits and co-added depth? Is there any benefit to real-time exposure time optimization to obtain nearly constant single-visit limiting depth?}\\No benefit from a constant single-visit depth.
    \item \textit{Are there any other potential trade-offs to consider when attempting to balance this proposal with others which may have similar but slightly different requests?}\\ Other proposals that are similar in science case or strategy definition:
\begin{itemize}
\item Bianco et al. with Explosive Physics \& Fast Transients LSST Cadence proposed a \textit{Presto-Colore} strategy which is in some aspects similar to ours - they suggest 3 visits per night, 2 in different filters with a short time interval between them ($\sim$ 1 hour), and another visit in one of those filters repeated after a longer time ($\sim$ 4 hours). There are no conflicts between their strategy and ours, although we would not require such frequent sampling, since TDEs are evolving much slower than fast transients.
\item Gezari et al. with An Extreme Rolling Cadence Wide-Fast Deep Survey suggest an extreme case of rolling cadence intended especially for TDE search, in which the whole survey footprint would be observed only int the first and in the last year of the survey, while in the rest of the time they would divide the sky into 8 strips of 2500 deg$^2$ each, observing one strip per year with preference to observations in \emph{u, g} and \emph{r} bands. While this would surely yield very well sampled TDE light curves in the pre-peak and peak phase (around 200 per year), the light curves observed in the first and the last year (around 1600 TDEs) will not have a better temporal sampling than in case of any other non-rolling cadence. Due to such monitoring of the sky we also estimate that roughly 2/3 of the events will be lost.
\end{itemize}
\end{enumerate}

\section{Performance Evaluation}
To evaluate the performance of a given survey strategy, we simulated
observed light curves of TDEs with LSST simulation framework. The 
details of our simulations can be found in Bricman\&Gomboc (in prep.).
Using different cadences released with the White Paper call, we
estimated the total number of discovered TDEs in 10 years of
LSST observations and the properties of simulated observed light curves
for each cadence.

Our simulations were done only on a small patch of the sky of
20.25 deg$^2$, which is covered in \texttt{GalaxyObj} base catalog
model in \texttt{CatSim}, containing around 17 million galaxies.
After simulating light curves of TDEs based on their rates and
other parameters, which define the light curve of a certain 
event, we estimated the number of discovered TDEs on 
the whole sky observed by the LSST, but avoiding the Galactic plane.
We do not expect TDEs from the Galactic plane, since the dust
absorbs the light from extra-galactic transients. Subtracting the 
area of the Milky Way from the total area of the sky observed by the 
LSST ($\sim$ 18000 deg$^2$) results in 13300 deg$^2$ of the sky 
from which we expect TDEs.

\begin{figure}
\centering
\includegraphics[width=\textwidth]{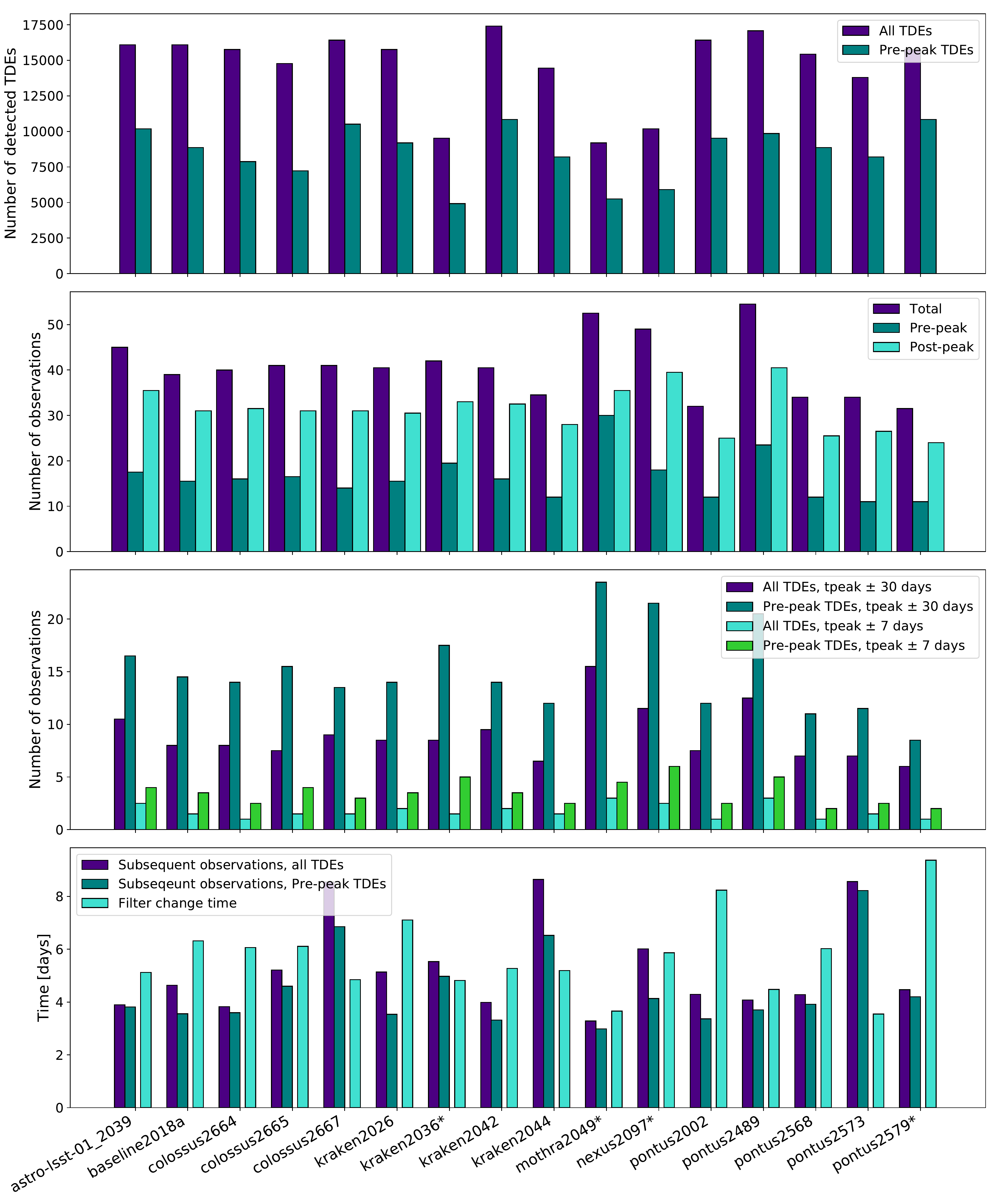}
\caption{\small{\bf Properties of detected TDEs}: in the upper panel the expected number of discovered TDEs with LSST in 10 years of observations on 13300 deg$^2$ is shown for all TDEs and TDEs discovered in pre-peak phase. The second panel shows average total number of observations, number of observations in pre-peak time, and number of observations in post-peak time. Third panel shows the number of observations in peak time $\pm$ 7 days, number of observations in peak time $\pm$ 30 days, both for all TDEs, and TDEs discovered in pre-peak phase. The fourth panel shows the time interval between subsequent observations for all and pre-peak TDEs, and the time interval in which filters are changed. Note that values are averaged over all detected TDEs and may largely vary between individual cases. Rolling cadences are denoted with ${}^*$.}
\label{fig:tdeNumbers}
\end{figure}

As good positive detections we choose all TDEs with 10 or more 
observations above the (limiting - 2) magnitude of the telescope in 
any band. Setting the constraint to (limiting - 2) magnitude excludes all
the light curves, in which observing points fall rather quickly
in the area close to the limiting magnitude. There the magnitude
error-bars are large, which would make the classification of
TDEs more difficult. The estimated numbers of detected TDEs
are shown in the first, top panel of Figure \ref{fig:tdeNumbers}. 
The numbers do not vary much from one cadence to another,
although in the case of rolling cadences, the number of discovered
TDEs tend to be smaller by a factor of $\sim$ 1.5.

However, the number of detected TDEs is not the only criteria
used to evaluate the performance of different cadences. The first
panel of Figure \ref{fig:tdeNumbers} also shows the number of TDEs
discovered in the pre-peak phase. In all cases of cadences, between
50\% and 60\% of detected TDEs will be discovered in pre-peak phase, 
with only one rolling cadence (\texttt{pontus\_2579}) having as much
as 68\% of TDEs observed in pre-peak phase.

In the second panel of Figure \ref{fig:tdeNumbers} the average number 
of observations of a detected TDE in total, in pre-peak and post-peak phases are shown. The
total average number of observations is of the same order of magnitude for all 
cadences, while the average number of pre-peak observations shows an
increase in case of rolling cadences and in case of \texttt{pontus\_2489} cadence,
which has more visits, but two 15 s exposures in a visit are replaced by
one 20 s exposure. It is not surprising that rolling cadences sample
the light curves better, also in the post-peak phase. 

From the third panel in Figure \ref{fig:tdeNumbers} it is obvious, that none of the
proposed cadences sample the light curves of TDEs in peak time $\pm$ 7
days very well. The same goes for observations in peak time $\pm$ 30 days. TDEs detected
in pre-peak phase are more frequently sampled in peak time $\pm$ 7 days and
in peak time $\pm$ 30 days.

The average time interval between two subsequent observations is shown in the
last panel of Figure \ref{fig:tdeNumbers} for all TDEs and TDEs discovered
in pre-peak phase, together with the time interval between two subsequent
filter changes. For all TDEs two observing points are on average between 
4 and 8 days apart, while for TDEs discovered in pre-peak phase, this time
is only slightly shorter. The filters are changed on average every 2-6 days.

We cannot claim that any of the proposed cadences would be
ideal for our science case. The concern with the rolling cadences is that
even though they seem to sample TDEs more frequently, especially in 
the pre-peak phase, they do tend to lose approximately half of the events 
that other cadences detect. On the other hand, for the case of 
non-rolling cadences the intervals between two subsequent observations 
and the time intervals between filter changes seem to long for
a successful classification of TDEs.

We propose a cadence with 2 visits in different filters per night every night
or every second night, which would sample TDE light curves
frequently enough. Due to the low number of discovered TDEs to date,
we cannot be certain whether 10 observing points we are assuming here as the minimum
will be enough for a positive TDE classification. That is why the filters
should be changed between subsequent visits, in order to calculate colors 
and color evolution, which are the basic parameters for the rapid TDE 
classification among other transients.

\vspace{.6in}

\section{Special Data Processing}
No special data processing beyond the standard LSST Data Management pipelines 
required.

\section{Acknowledgements}
KB and AG acknowledge the financial support from the Slovenian Research Agency.
This work was developed partly within the TVS Science Collaboration and the authors acknowledge the support of TVS in the preparation of this paper.

\section{References}
\renewcommand{\bibsection}{}
\bibliographystyle{aasjournal}
\bibliography{references}{}

\end{document}